# Tuning magnetic frustration of nanomagnets in triangular-lattice geometry

X. Ke, J. Li, S. Zhang, C. Nisoli, V. Crespi, and P. Schiffer*

*Department of Physics and Materials Research Institute, Pennsylvania State University,*

*University Park, PA 16802, USA*

We study the configuration of magnetic moments on triangular lattices of single-domain ferromagnetic islands, examining the consequences of magnetostatic interactions in this frustrated geometry. By varying the island-island distance along one direction, we are able to tune the ratio of different interactions between neighboring islands, resulting in a corresponding variation in the local correlations between the island moments. Unlike other artificial frustrated magnets, this lattice geometry displays regions of ordered moment orientation, possibly resulting from a higher degree of anisotropy leading to a reduced level of frustration.

*schiffer@phys.psu.edu

The two-dimensional triangular lattice of spins is perhaps the simplest structure in which nearest neighbor spin-spin interactions can be frustrated [1]. Ideal Heisenberg spins with nearest neighbor antiferromagnetic interactions on such a lattice are expected to have an ordered 120° spin configuration. However, antiferromagnetically coupled Ising spins sharing the same anisotropy axis are strongly frustrated and are expected to display a disordered magnetic ground state [2]. With further-neighbor or anisotropic interactions, the ground state degeneracy of such Ising spins is greatly reduced, and a long-range ordered ground state is expected [3,4].

While numerous compounds include magnetic ions residing on a triangular lattice [5], the individual spins in such materials cannot easily be probed to determine the local arrangement of moments and how they accommodate the magnetic frustration. We report studies of a triangular system in which the local moments can be examined directly: lattices of nanometer-scale single-domain ferromagnetic islands. By manipulating the distance between islands, we are able to tune the inter-island magnetostatic interactions. After demagnetization to minimize the magnetostatic energy, the resulting moment configuration reveals the formation of regions of regular ordering–reflecting a lower level of frustration in this system than is seen in other artificial frustrated magnets.

Our studies follow previous work on frustrated arrays of nanometer-scale ferromagnetic elements: both square arrays of single-domain ferromagnetic islands with perpendicular nearest-neighbors [ 6 , 7 ] and hexagonal networks of single-domain ferromagnetic nanowire links or islands [8,9,10]. Our triangular lattices are formed of stadium-shaped permalloy islands (shown schematically in Fig. 1) with lateral dimensions of 220 nm x 80 nm and a 25 nm thickness [6,7]. These islands were patterned by e-beam lithography followed by molecular beam epitaxy deposition and lift-off. The horizontal inter-island distance (center-to-center) is fixed at 400 nm while the vertical distance ($2d$) between the horizontal rows varies from 320 nm to 1760 nm. Each island forms a single ferromagnetic domain with its moment directed along the island's long axis, due to shape anisotropy. Thus the islands can effectively be treated as giant Ising spins. We calculate the inter-island interaction energy ($J$) using finite size micromagnetic methods [11]. The results are shown in Fig. 2 for antiparallel spin alignment of the island pairs. $J_1$, $J_2$, and $J_3$ are the interaction between two horizontal, diagonal, and vertical nearest-neighbor islands, respectively. $J_1$ is a constant, while the interactions $J_2$ and $J_3$ depend strongly on $d$. As expected, $J_1$ is negative, and $J_3$ is positive and decreases with increasing $d$. Interestingly, $J_2$ changes sign from negative to positive with increasing $d$, a property that is reflected in the measured moment configuration discussed below.

To minimize the magnetostatic energy of the arrays, we followed our previously developed AC demagnetization procedure of rotating the arrays at 1000 rpm in a stepwise decreasing in-plane magnetic field [6,7,12]. The field was reduced from 2000 Oe (well above $H_c \sim 770$ Oe of the islands [7]) to 0 by constant size steps of 3.2 Oe in field magnitude, reversing the field direction at each step. After demagnetization, magnetic force microscopy (MFM) images of 400-750 islands were acquired at 4 different locations far from the edges of each array. The uncertainty of quantities derived from the images is estimated as the standard deviation of the individual image results.

Figure 3 shows MFM images of moments in arrays with $d = 160$ nm, 320 nm, 480 nm, and 880 nm after AC demagnetization (the right-hand inset to Fig. 1 shows a magnified portion of the $d = 400$ nm lattice in which the individual island moments are clearly seen). The black and white dots correspond to the south and north poles of the island moments, respectively. These images demonstrate our ability to resolve the moment configurations after demagnetization, and analysis of the full MFM images demonstrates that the arrays have zero residual moment within statistical error after the AC demagnetization. One striking feature of these images is the prevalence of distinct regions in which the moments are ordered in regular patterns, as outlined in color in Fig. 3 and discussed in more detail below.

To quantitatively characterize the demagnetized magnetic state of our arrays, we consider individual triangles within the lattice structure. We categorize different arrangements of the moments on a triangle as one of three possible moment configurations (Types I, II, and III) shown in Fig. 4a. In Fig. 4b we plot the fractional distribution of these different moment configurations as a function of $d$, counting all triangles. We combine symmetry-equivalent triangles pointing either up or down with inverted moments into the same category. For the smallest value of $d = 160$ nm, $J_3$ is dominant, and $J_1$ and $J_2$ are roughly equal. The balance of $J_1$ and $J_2$ corresponds to an antiferromagnetic triangular Ising model, where each triangular plaquette is expected to be frustrated, with either Type II or Type III triangles. However, the distribution of triangle moment configurations for this value of $d$ is close to that expected for a random population of islands (approximately 25% of both Type I and III triangles and 50% of Type II triangles), presumably due to the strong influence of $J_3$. With increasing $d$, $J_3$ decreases and $J_2$ changes sign. Correspondingly, more diagonally neighboring island pairs have parallel moment alignment, which increases the fraction of Type I and decreases that of Type III triangles. The fraction of Type I triangles is maximized near $d$ = 320 nm, where $J_2$ is also maximized (Fig. 2). Indeed, this situation ($J_2 > |J_1| > J_3$) corresponds to a square ferromagnetic Ising system with a smaller antiferromagnetic perturbing next-nearest-neighbor coupling ($J_1$), a model extensively studied and expected to exhibit a polarized ground state of Type I triangles [13]: large ordered regions of Type I are indeed seen in Fig 3b. Since the AC demagnetization prevents a condensed polarized phase, regions of opposite orientations can be observed. As $d$ further increases, $J_1$ remains constant while both $J_2$ and $J_3$ decrease and approach zero. The resulting dominance of inter-island correlations along the rows of islands is reflected in the preponderance of Type II triangles at our maximum values of $d$.

The above analysis of triangle statistics does not directly reflect the presence of regions in which there is regular ordering, as outlined in Fig. 3. The ordered regions of Type I and Type III triangles are visible in the image of the $d$ = 160 nm lattice, and the Type I regions dominate in the image of the $d$ = 320 nm lattice.  This corresponds well to the fractional population of the different triangle types which is expected from the energy arguments discussed above.  As seen in Fig. 3d, for our largest value of $d$ = 880 nm, the triangular arrays are best described as isolated chains, and the moments along the chain tend to have anti-parallel spin alignment; sections of the chains in which moments have this alignment are outlined in Fig. 3d.

The regions of ordered moments typically span 5-15 lattice spacings in their largest dimensions and 1-5 lattice spacings in their smallest dimensions (a size which appears unchanged by demagnetization with larger 16 Oe steps), and thus correspond to only short-range ordering. While our MFM scans cover several hundred islands, the technique does not allow very large-scale scans that would be needed to collect exhaustive statistics on the geometries of these intermediate-scale ordered regions. Additionally, because of their finite size, statistical studies of the average longer-range correlations between island pairs only show correlations which can be derived directly from the three types of nearest neighbor correlations (corresponding to $J_1$, $J_2$, and $J_3$) [14]. On the other hand, no such short-range ordered regions are observed in frustrated square arrays [6,7,12] or hexagonal networks [9,10]. Furthermore, our observation of short-range ordered regions is quite distinct from the fully disordered magnetic state expected for an ideal Ising triangular antiferromagnet with nearest-neighboring interactions [2]. We attribute the short-range ordering to the nonequivalent interaction strength of island pairs along different directions (e.g., the anisotropy intrinsic to the triangular lattice) and the interactions with further neighboring islands [15], which result in weaker magnetic frustration than in other artificial frustrated systems [6,9,10,12].

Our results have implications both for nanomagnet arrays as well as for the larger class of frustrated magnets. For nanomagnets arrays, our data demonstrate ordering which has not previously been observed in frustrated geometries, but which is notably finite in its range. Future studies will be needed to probe why the regions are limited in size and what factors (demagnetization protocol, lattice or island geometry, interaction strengths, etc.) control their extent. More broadly, the existence of short-range-ordered regions of spins in frustrated systems leading to so-called cluster glass states [16,17] has been well-documented, and this model system may be able to lend insight into the nature of such states and how they form.

We acknowledge the financial support from Army Research Office and the National Science Foundation MRSEC program (DMR-0820404) and the National Nanotechnology Infrastructure Network. We are grateful to Prof. Chris Leighton for the film deposition.

**FIGURE CAPTIONS**

Figure 1. (a) Schematic of the triangular lattice under study. The nearest neighbor island-island distance is fixed at 400 nm along the horizontal direction while the row separation, $2d$, along vertical direction is varied. The inset on the left shows a typical AFM image of islands with $d = 400$ nm, and the inset on right shows a MFM image where the black and white dots correspond to the magnetic poles.

Figure 2. Calculated dipolar interaction $J_1$, $J_2$, and $J_3$ for the nearest neighbor island pairs along horizontal, diagonal, and vertical directions, respectively, assuming that the island pairs have anti-parallel spin alignment.

Figure 3. Typical MFM image for the array with $d = 160$ nm (a), 320 nm (b), 480 nm (c) and 880 nm (d). Red loops indicate regions of ordered type I configurations, green loops type II, and yellows loops type III. The scale bars are 2 µm. The green loops in (d) indicate regions in which an alternating alignment of moments is seen.

Figure 4. (a) Schematic of the 3 different types of triangle moment configurations. (b) The fractional population of each type, as a function of row spacing.

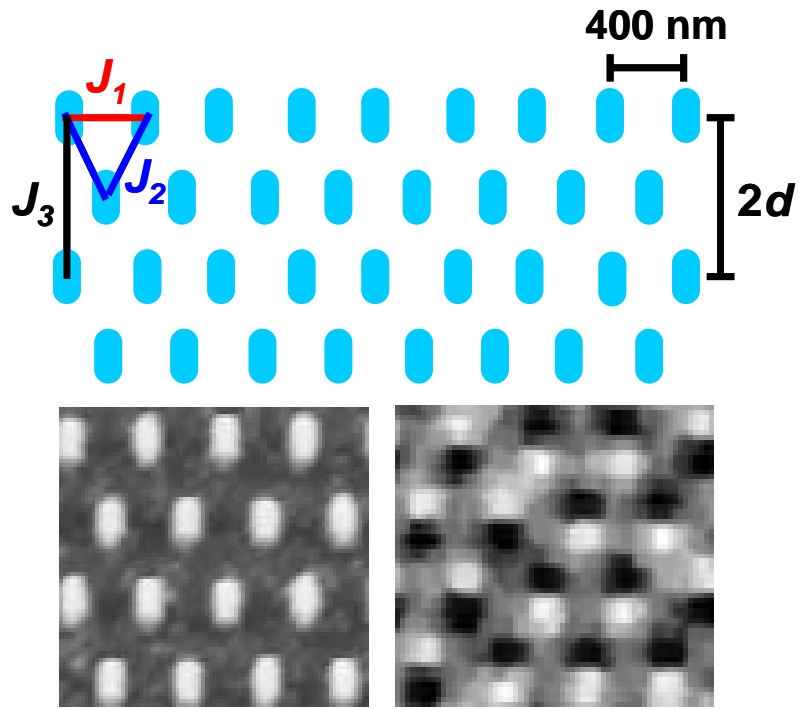

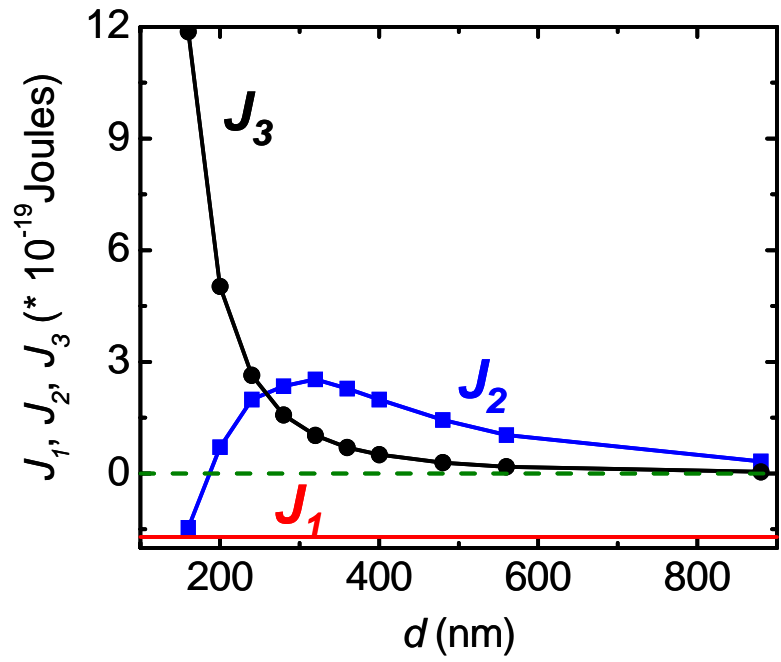

Figure 3.
X. Ke *et al.*

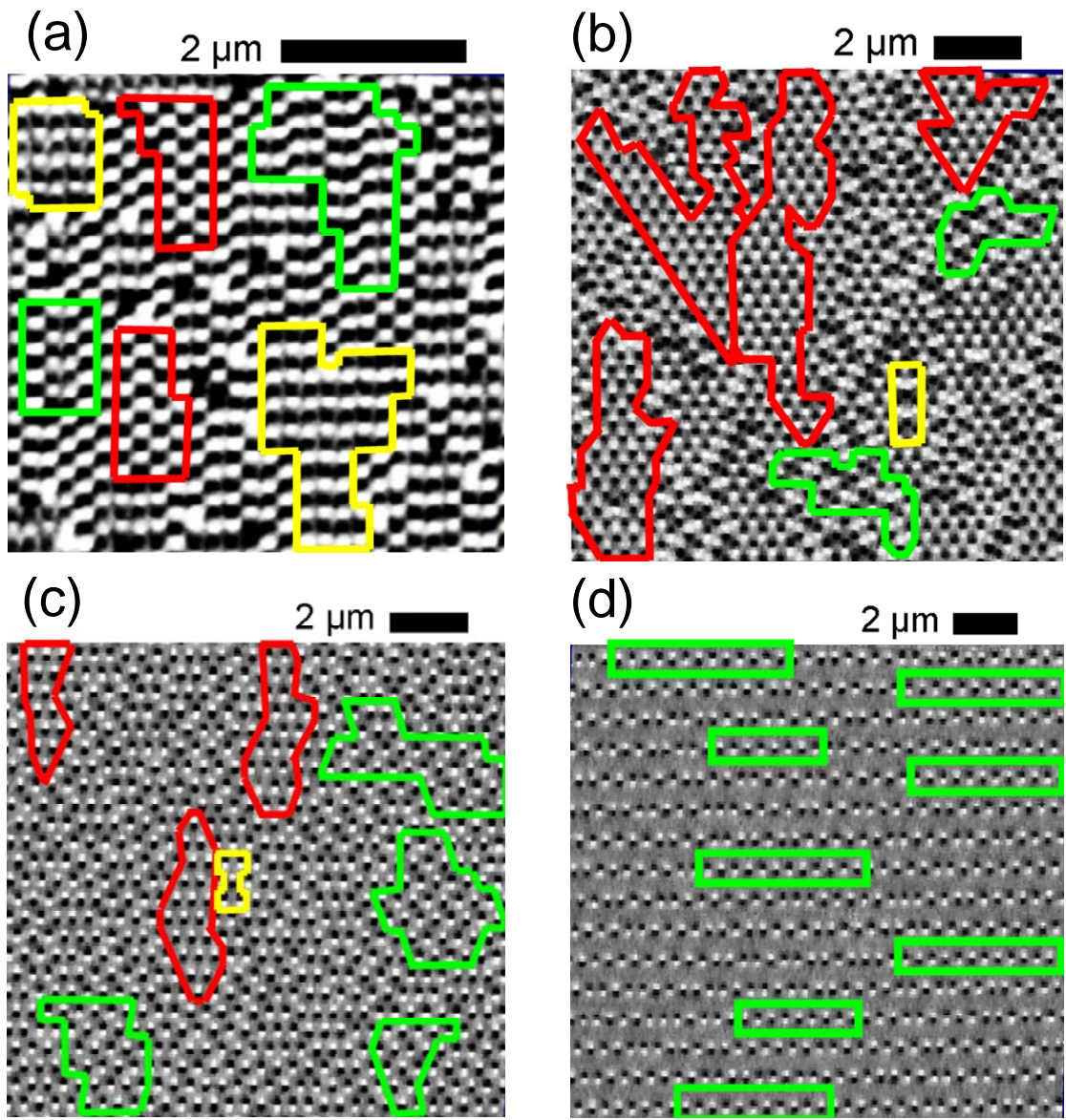

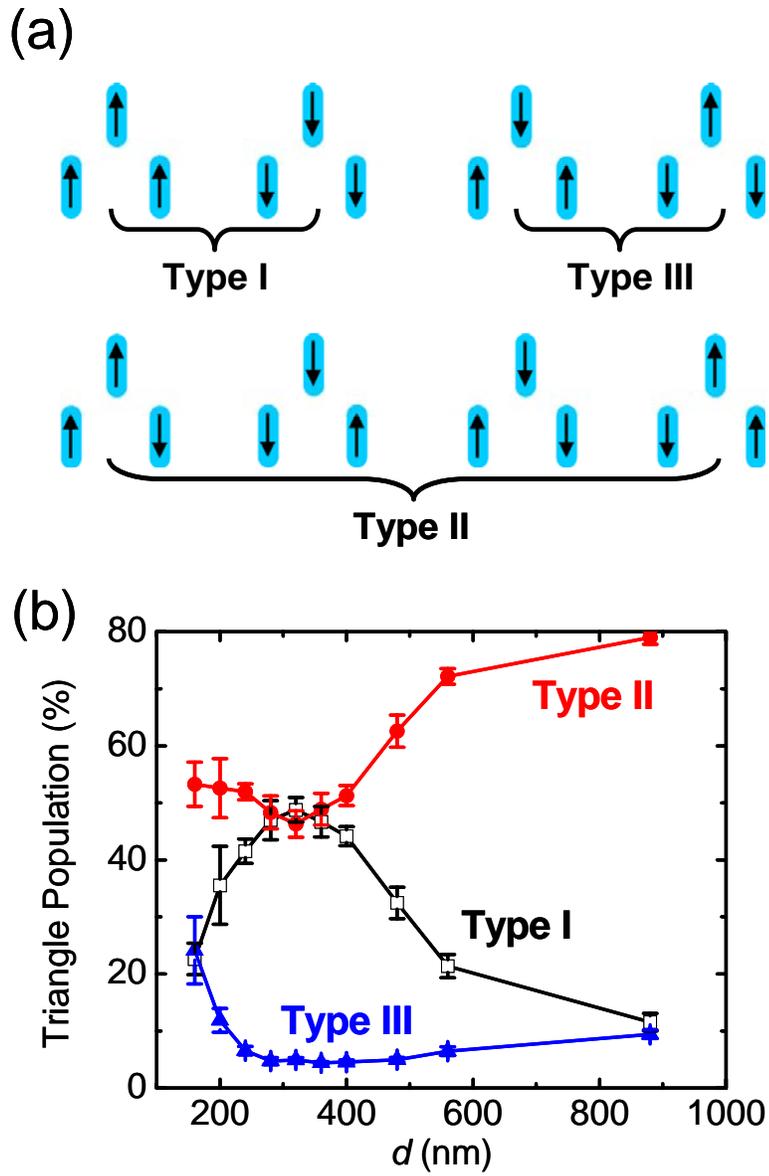